\documentclass[lettersize,journal]{IEEEtran}
\usepackage{cite}
\usepackage{amsmath,amssymb,amsfonts}
\usepackage{algorithmic}
\usepackage{graphicx}
\usepackage{textcomp}
\usepackage{xcolor}

\usepackage{braket}
\usepackage{physics}
\DeclareMathOperator\erfc{erfc}
\usepackage{nccmath}

\def\BibTeX{{\rm B\kern-.05em{\sc i\kern-.025em b}\kern-.08em
    T\kern-.1667em\lower.7ex\hbox{E}\kern-.125emX}}
\usepackage{balance}

\begin{document}
\title{Quantum Keyless Private Communication 
with Decoy States for Space Channels
}
\author{Ángeles Vázquez-Castro, \textit{IEEE Senior Member}, Andreas Winter, Hugo Zbinden
\thanks{
Ángeles Vázquez-Castro
is with the Department of Telecommunications and Systems Engineering, and with The Centre for Space Research (CERES) of Institut d'Estudis Espacials de Catalunya (IEEC-UAB) at 
Autonomous University of Barcelona,
Barcelona, Spain.
Email: angeles.vazquez@uab.es.}
\thanks{
Andreas Winter is with ICREA and 
 with the Quantum Information Group (GIQ) at the Physics 
 Department, Autonomous University of Barcelona, Barcelona, 
 Spain. Furthermore with the Institute for Advanced Study, 
 Technical University Munich, Garching, Germany. Email: 
 andreas.winter@uab.cat.}
\thanks{
Hugo Zbinden is an Honorary Professor with the Department of Physics at 
University of Geneva and Group Leader at Quantum Communication Technology Group in Vigo Quantum Communication Center, Spain.
Email: hzbinden@vqcc.uvigo.es.}}


\maketitle

\begin{abstract}
With the increasing demand for secure communication in optical space networks, it is essential to develop physical-layer scalable security solutions. In this context, we present the asymptotic security analysis of a keyless quantum private communication protocol that transmits classical information over quantum states. Different from the previous literature, our protocol sends dummy (decoy) states optimally obtained from the true information to deceive the eavesdropper. We analyze optical on-off keying (OOK) and binary phase shift keying (BPSK) for several detection scenarios. Our protocol significantly improves the protocol without decoy states whenever Bob is at a technological disadvantage with respect to Eve. Our protocol guarantees positive secrecy capacity when the eavesdropper gathers up to 90-99.9\% (depending on the detection scenario) of the photon energy that Bob detects, even when Eve is only limited by the laws of quantum mechanics.

We apply our results to the design of an optical inter-satellite link (ISL) study case with pointing losses, and introduce a new design
methodology whereby the link margin is guaranteed to be secure by our protocol. Hence, our design does not require knowing the eavesdropper’s location and/or channel state: the protocol aborts whenever the
channel drops below the secured margin. Our protocol can be implemented with state-of-the-art space-proof technology. Finally, we also show the potential secrecy advantage when using (not yet available) squeezed quantum states technology.
\end{abstract}

\begin{IEEEkeywords}
Quantum channel, wiretap channnel, secret communication, decoy state.
\end{IEEEkeywords}

\section{Introduction}
\subsection{Motivation and intuition of our protocol}
Free-space optical communication is a promising technology for improving the connectivity of space networks. Its potential applications range from mega-constellations to terrestrial and non-terrestrial integration for 6G. However, despite the high directionality of laser beams, optical communication is still exposed to the threat of eavesdropping, particularly due to pointing errors. Hence, there is a need for the development of keyless physical-layer security protocols that are scalable and impervious to the rapidly escalating computational power of quantum computers, e.g. by taking advantage of the quantum properties of light. The original idea for these protocols, known as the wiretap channel, was first proposed by Wyner in 1975 \cite{Wyner1975}. The basic idea is elegant: it exploits the inherent randomness of the physical communication channel as an entropy source to prevent leakage of information towards potential eavesdroppers. The main strengths are essentially two. First, it guarantees information-theoretic post-quantum security, i.e. no limitation is assumed on the adversary's computational power. Second, it is a keyless protocol, which makes it a scalable protocol thus attractive for the many upcoming massive connectivity scenarios.

 The idea was generalised by Csiszár and Körner in \cite{Csiszar1978} who characterized the secrecy capacity for the general discrete memoryless wiretap channel and it was further strengthened to meet cryptographic security standards in \cite{Maurer1993,Bennett1984,Hayashi2006,Hayashi2011,Bellare2012}. A number of comprehensive surveys and tutorial papers are available on the principles of the wiretap channel and the different operational applications, which include not only confidentiality (secrecy, our focus on this work) but also authentication, integrity and key generation \cite{Bloch2011,Oggier2021,Mahdi2021}. While Wyner's approach 
shows that secret communication is guaranteed if the mutual information to the eavesdropper (Eve), $I_E$, is smaller than the mutual information to the legitimate receiver (Bob), $I_B$ (with a certain input conditional distribution), the practical implications of the underlying (classical and quantum) information-theoretic results go well beyond  $I_B - I_E$. For example, \cite{Khisti2007} makes use of the results by Csiszár and Körner \cite{Csiszar1978} and propose an "artificial noise" based scheme whereby the transmitter sends information along the directions corresponding to non-zero singular values of the legitimate multi-antenna channel, while transmitting artificial noise in its null space. These ideas were later followed up for more scenarios \cite{Goel2008,Zhou2010,Xiong2012,Ghogho2012,Sikri2021}. As another example for satellite links, the seminal results by Maurer \cite{Maurer1993} were followed up in \cite{Vazquez2018,Two,Hayashi2020} to develop practical two-way protocols that guarantee $I_B - I_E$ to be always positive irrespective of Eve's location and/or channel. 

In this work we also make use of results obtained by Csiszár and Körner in \cite{Csiszar1978} and investigate their impact on secure direct communication over quantum semi-classical and non-classical quantum states. Specifically, let's denote $X$, $Y$ and $Z$ as the (classical) random variables of the legitimate transmitter Alice, legitimate receiver Bob, and Eve, respectively. Instead of considering the usual Markov chain $X \rightarrow Y \rightarrow Z$ that leads to the condition $I_B - I_E$ we consider the Markov chain $V \rightarrow X \rightarrow YZ$. In both cases, the Markov chains induce the \textbf{channel stochastic degradation conditions} that define the entropic properties of the channel. Specifically, the entropic content of $Z$ depends in our case on the quantum properties of the received quantum systems by Bob and Eve. The intuition of our practical implementation of the auxiliary variable $V$ is that it represents optimized "dummy" optical pulses that help to exponentially decrease the amount of information received by Eve in the finite-length regime. 
\subsection{State of the art and main contributions}

Our focus is free-space optical (FSO) communication over quantum satellite channels when classical information is carried over (semi-classical or non-classical) quantum states. Despite the high directionality of a laser beam, this channel can still be wiretapped for example when the (ideally) Gaussian beam width of a laser is wider than the receiver size, which is a typical case for satellite communications \cite{Agaskar2013}. While the secrecy performance for classical terrestrial FSO communication has been widely analyzed, literature is rather scarce for quantum communications. This is due to the fact that quantum key distribution (QKD) is currently deemed as quantum communications. However, in this paper we follow Shannon theoretic terminology and the corresponding 
 mathematical framework \cite{Shannon1948,Shannon1949} (in particular, note that QKD does not does not perform quantum communication à la Shannon, but it distributes quantum randomness for Alice and Bob to distill a shared secret key \textit{after} perfect secret communication can take place via e.g. one-time-pad \cite{Shannon1949}). 

For the \textbf{radio frequency case} assuming the channel degradation condition and one-way wiretap protocol, the time-varying mobile channel has been widely analyzed \cite{Liang2008,Liang2009,Bloch2011,Wu2018}. However, the channel degradation condition is quite impractical at radiofrequency as it is rather unrealistic to either make assumptions on Eve’s location or have feedback for full or partial channel side information (CSI) (while it can still be feasible in specific scenarios \cite{Chorti2022}). Our paper \cite{Vazquez2018} considered this problem for the satellite Gaussian channel. Unfortunately, the degradation condition is also too strong in this scenario because if the wiretapper eavesdrops the information by being located in between the sender and the legitimate sender (passive man-in-the-middle attack), information-theoretic secure communication cannot be guaranteed. To resolve this problem, we introduced a two-round protocol inspired by Maurer \cite{Maurer1993} that  guarantees secure communication even against the passive man-in-the-middle attack for the satellite Gaussian channel case \cite{Two}. We also proposed it for the optical Poissonian channel as discussed next.

For the \textbf{optical case} assuming the channel degradation condition and one-way wiretap protocol, the work \cite{Martinez2015} derives and analyzes secrecy assuming the optical channel is a Gaussian channel impaired by the Beer-Lambert attenuation model, which is a very different channel model to the space channel models. Our two-round protocol resolving the passive man-in-the-middle attack for the satellite Poissonian channel case for space scenarios is presented in \cite{Hayashi2020} and ensures $I_B - I_E$ to be always positive irrespectively of Eve's location and/or channel state (at the cost of higher delay and complexity). Similar ideas are presented in more recent works such as \cite{Saber2018} for the multiple-input multiple-output multi-apertures channel, where the authors show that adding additional apertures can improve the performance. A related publication to the work we present here is \cite{Endo2015}, where the authors show that information-theoretic security from a one-way wiretap protocol under the channel degradation condition outperforms QKD in distance, enabling secure optical links between geostationary Earth orbit satellites and ground stations. The authors also consider the Markov chain $V \rightarrow X \rightarrow YZ$ as we do here, and observe that the result of the preprocessing map $V - X$ has a certain analogy with the decoy QKD protocol \cite{Lo2005}. However, this work only assumes on-off Keying (OOK) modulation and Poissonian statistics.

Assuming the channel degradation condition, our previous work \cite{Angeles2021} proposes a one-way wiretap protocol with OOK over coherent states showing high-security rates for the space-to-ground satellite channel when Eve is only limited by the laws of quantum mechanics. However, this work still makes assumptions about Eve's channel, which we remove in the new protocol we present here. While the wiretap principle can be used for key distillation (see e.g. \cite{Chorti2022} \cite{Fujiwara2018,DW2005,Pan2020}) in which case it is to be compared with QKD, such application is out of our scope. On the other hand, there is some work investigating the consequences of restricted eavesdropping for Eve on the performance of QKD links \cite{Shapiro2019,Shapiro2019-1,Ling2019}. It is worth noting that the additional assumptions made by the QKD community in these investigations are very similar to the assumptions for the wiretap-based protocols. Finally, our protocol is also comparable to quantum secure direct communication (QSDC) protocols, which, like our protocol, enable secure transmission of messages directly from the sender, Alice, to the receiver, Bob  (see e.g. \cite{DPan2023} for a comparison between QSDC and our previous protocol \cite{Angeles2021}).

\begin{figure*}[!ht]
    \centering
    \includegraphics[width=\textwidth]{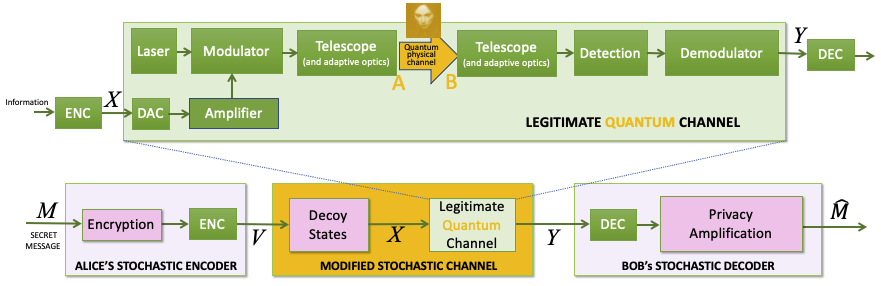}
    \caption{Logic flow of the proposed quantum keyless secure communication protocol over an (authenticated) quantum channel using a weak pulses laser. Top: legitimate quantum optical channel (i.e. without any security measure) showing the quantum system at transmission by Alice, A, and at reception by Bob, B. Bottom: legitimate quantum optical channel enhanced with our security functionalities: (keyless) encryption module before channel encoding, addition of decoy states and decoding and privacy amplification at Bob.}
    \label{fig:SysModel}
\end{figure*}
Our fundamental contribution is two-fold. First, our protocol departs from the wiretap related literature, by not requiring any assumptions about Eve's channel. This is achieved thanks to two main components in our protocol design. The first one is the decoy states we optimally introduce in our protocol, for which we compute and analyze the secrecy capacities, which shows positive secrecy rates (i.e. it guarantees information-theoretic security) even if Eve gathers 99.9\% of the photon energy Bob gathers. The second one is that we incorporate an "abort" feature (like QKD protocols) by defining a "secure link margin", which is a security version of the well-known "link margin" of satellite communications design. Within such margin, our protocol guarantees reliability and informational-theoreticly security. Hence, by monitoring the signal-to-noise ratio (SNR) at the reception, whenever it drops below such margin, the protocol aborts. Our second contribution is the practical implementation of our protocol by working out a realistic and relevant use-case (inter-satellite links), which can be implemented in space with state-of-the-art space-proof equipment. We remark that our use of the term "decoy state" is purely semantic and does not refer to any specific method used in QKD protocols (while it expands the conceptual use of the term beyond the QKD framework).
In the following, Sections II and III present the protocol and models, IV and V present our secrecy capacity derivations for three detection different scenarios, VI presents a practical design based on a secure link margin for the optical inter-satellite link (ISL) use-case while VII shows the potential advantage of using squeezing states instead of coherent states. Finally, Section VIII discusses conclusions and further work. 
\begin{figure}[tbh]
    \centering
    \includegraphics[scale=0.2]{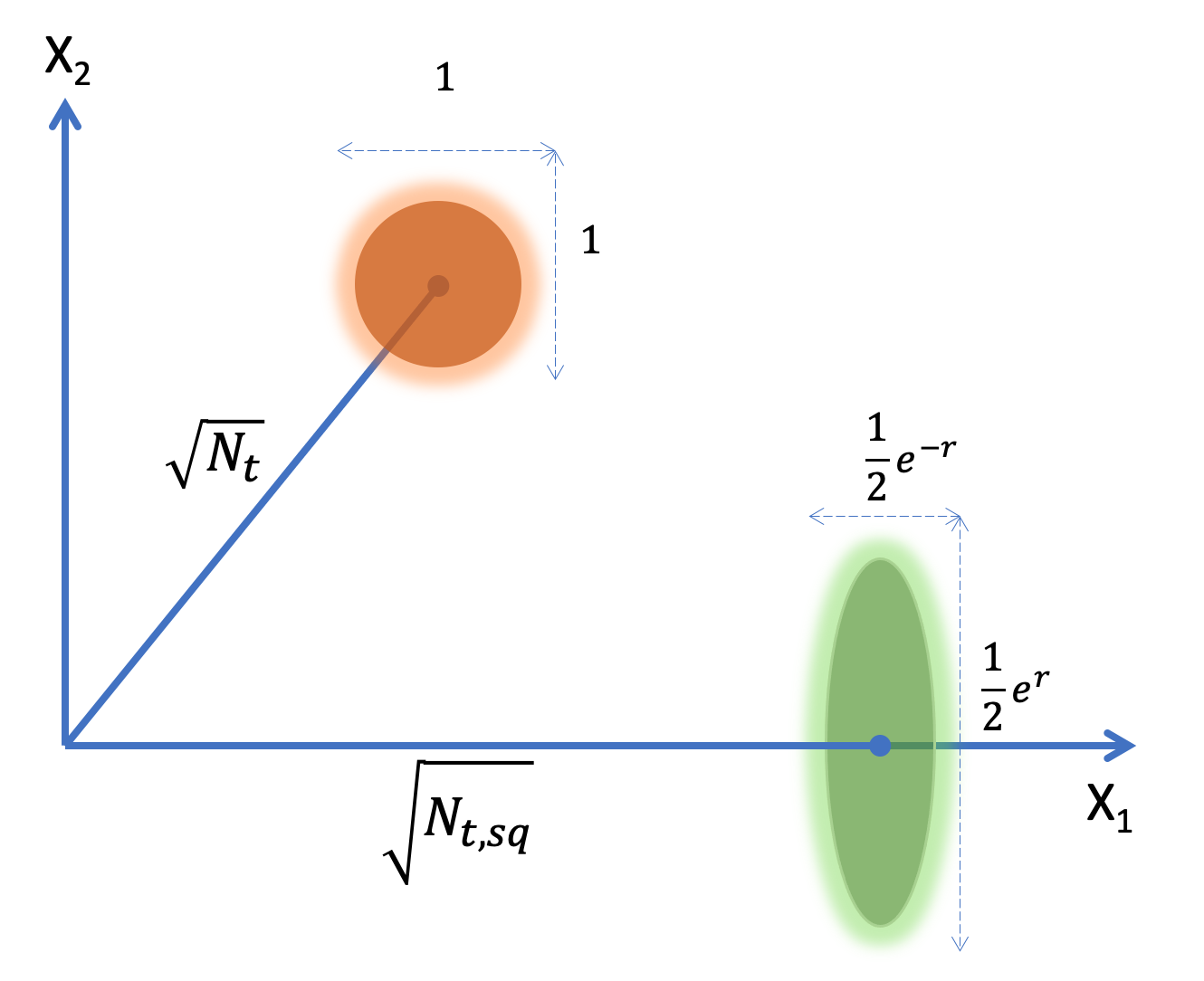}
    \caption{Illustration of a quantum (semi-classical) coherent state (orange) and of a quantum (non-classical) squeezed coherent state (green), showing the variance of the respective uncertainties.}
    \label{fig:CohSqz}
\end{figure}

\section{Quantum Keyless Private Communication Protocol and signal models}
\subsection{Quantum Keyless Private Communication Protocol}
A standard optical communication channel without any security measure is shown in Fig. 1 (top). Alice sends a stream of information bits to a channel encoder, which outputs the encoded information represented by the random variable $X.$ For each use of an (authenticated) channel, the transmitter prepares a quantum state modulated by the random variable $X \in \mathcal{X}=\{0,1\}$ (see Fig. \ref{fig:SysModel}), with input probability denoted as $q_x$. In this work we mainly focus on the practical case of coherent quantum states, however, we will also discuss the potential gains when using squeezed states (see Fig. \ref{fig:CohSqz}). Such a channel is threatened by Eve who has access to it and thus the information carried by the quantum state is not secure.

From our Shannon framework perspective, this means the information obtained by Bob, represented by the random variable $Y$, is received with guaranteed reliability but not with guaranteed security. Our method adds a few functionalities as shown in Fig. \ref{fig:SysModel} (bottom) to also guarantee security in addition to reliability.

\section{information-theoretic preliminaries and detection scenarios}
\subsection{Secrecy capacity}
A given wiretap channel $X \rightarrow YZ$ defines a 
pair of point-to-point channels, one for Bob's and one Eve's 
output, $X \rightarrow Y$ and $X \rightarrow Z$, respectively. 
We refer to these two channels as the ``main channel'' and the 
``eavesdropper's channel''. Secret wiretap coding is possible 
because the former is ``better'' in some sense than the latter, 
and impossible due to it being similarly ``worse'' \cite{Watanabe2014}. In this section, we first establish some relevant terminology of such ordering. Let's define the following differentiable function in the input probability, $P_{x}$
\begin{equation}
f(P_x) = I_B - I_E,
\end{equation}
where $I_B=I(X:Y)$ and $I_E=I(X:Z)$ denote the mutual informations of Bob and Eve about $X$, respectively, for the joint distribution 
of $X,\,Y,\,Z$ defined by $P_x$ and the channel $X \rightarrow YZ$. We say that the main channel is \textbf{more capable} than the 
eavesdropper's channel if $f(Px)>0$ for all $P_x \in \mathcal{S}$, with $\cal{S}$ defined as the simplex 
\[
\mathcal{S} = \left\{(p_x : x\in\mathcal{X}) : 
                     \forall x\ p_x\geq 0,\ \sum_x p_x=1 \right\}. 
\]
A more stringent relation is called  \textbf{less noisy} , and it is defined by the condition $I(V:Y) \geq I(V:Z)$ for all distributions $P_{VX}$ and Markov chains $V \rightarrow X \rightarrow YZ$. For a general wiretap channel $X\rightarrow YZ$, Csiszár and Körner characterized the secrecy capacity as \cite{Csiszar1978}
\begin{align}\label{main_opt}
C_s = \max_{V}  I(V:Y) - I(V:Z),
\end{align}
where the maximum is is over all random variables $V$ such that $V \rightarrow X \rightarrow YZ$ is a Markov chain. Without loss of generality, $|\mathcal{V}| \leq |\mathcal{X}|$. The cardinality bounds on the alphabets of the auxiliary random variable for the maximization to be computable is $|\mathcal{V}| \leq |\mathcal{X}|$. We note that 
\[
I(V:Y) - I(V:Z) = f(P_x) - [I(X:Y|V) - I(X:Z|V)],
\]
\noindent which shows that  $f(P_x)$ is not the maximum secrecy rate in general.

Hence, in our design we need to verify whether or not the quantum wiretap channels we are considering are more capable for different assumptions on Bob's and Eve's detection capabilities. Note that even if our physical channel is not classical as assumed in \cite{Csiszar1978}, here we assume practical optical technologies to gather the photon energy as shown in Fig. 1 (bottom), and therefore the random variables that we assume are classical. Moreover, the maximization \eqref{main_opt} is difficult to evaluate, and a simpler expression for the secrecy capacity is not known even for the simple cases when Bob's and Eve's channels are both symmetric. Hence, we impose the additional constraint of stochastic degradability. 

We say that Eve's channel is stochastically degraded with respect to Bob's channel when the following condition holds
\begin{align}\label{degradation}
P(z|x) = \sum_{y\in\mathcal{Y}} Q(z|y)P(y|x),
\end{align}
with a channel $Q:\mathcal{Y} \rightarrow \mathcal{Z}$. This implies that 
the main channel is more capable, and indeed less noisy that 
the eavesdropper's channel, but the conditions are not equivalent. Leung-Yan-Cheong \cite{Leung1977} showed that in the simple case when Bob's and Eve's channels are both symmetric, the secrecy capacity has a simple form as follows. Let us parameterize the stochastic degradation by the physical degradation parameter $\gamma$ introduced in the previous section, then, we have

\begin{align}\label{Cs_BSCs}
C_s &= \max_{V}  I(V:Y) - I(V:Z | \gamma),\\
&= \max_{X}  I(X:Y) - I(X:Z | \gamma),\\
&= h(P) -  h(P(\gamma)),\label{Cs_BSCs}
\end{align}
where $h(\cdot)$ is the classical entropy of a binary source and $P$ and $P(\gamma)$ denote the channel transition probabilities of Bob's and Eve's channels, respectively.
\subsection{Stochastic degradation condition}
Bob and Eve's channels are denoted as $W_B(\epsilon_{00},\epsilon_{01})$, and $W_E(\epsilon_{00}(\gamma),\epsilon_{01}(\gamma))$, respectively, (see Fig. 3) where the latter will be referred to as $W_E(\gamma)$ for short. These probabilities depend on the detection scenarios, which we will define next. However, the condition for stochastically degraded channel for a general binary channel $(W_B, W_E(\gamma))$ can be known \cite{Hayashi2020} as the stochastic degradation condition for this channel is equivalent to the condition that $(\epsilon_{00}(\gamma),\epsilon_{01}(\gamma))$ belongs to the quadrangle spanned by $(0,0)$, $(\epsilon_{00},\epsilon_{01})$, $(1,1)$ and $(1-\epsilon_{00},1)$. In general, we assume that $\epsilon_{00} \geq \epsilon_{01}$ and $\epsilon_{00}(\gamma) \geq \epsilon_{01}(\gamma)$, in which case the \textbf{stochastic degradation condition} is equivalent to 
\begin{align}
\frac{\epsilon_{01}(\gamma)}{\epsilon_{00}(\gamma)}\geq\frac{\epsilon_{01}}{\epsilon_{00}}
\end{align}
when  $\epsilon_{00}(\gamma) \leq \epsilon_{00}$ and 
\begin{align}
\frac{1-\epsilon_{01}(\gamma)}{1-\epsilon_{00}(\gamma)}\geq\frac{1-\epsilon_{01}}{1-\epsilon_{00}}
\end{align}
when $\epsilon_{00}(\gamma) \geq \epsilon_{00}$.

\begin{figure}[tbh]
\centering
\includegraphics[scale=0.3]{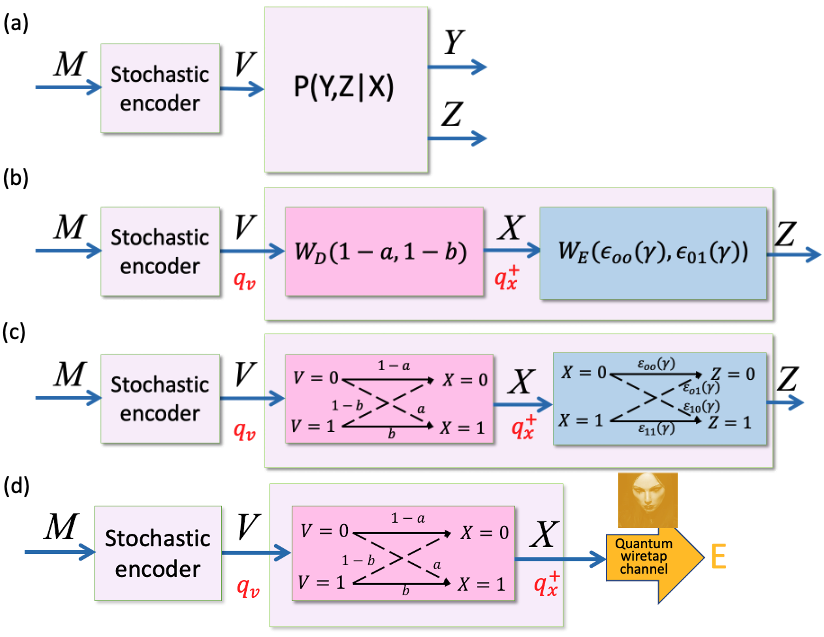}
\protect\caption{(a): General wiretap channel block diagram as proposed in \cite{Csiszar1978}. (b): Eve's general cascade channel when the channel degradation condition does not hold (wiretap channel is less noisy). (c): Same as (b) for our binary case and with our notation. (d): Binary wiretap channel model with our notation when Eve is only limited by quantum mechanics.} 
\label{fig:EveModel}
\end{figure}

In case the degradation condition does not hold, the generation of decoy states can be modeled as an additional channel in cascade with Bob and Eve's channels which not only modifies the statistical wiretap channel properties but also the average transmitted photon energy. We denote $p_{x|v}(0|0) = 1 - a$ and $p_{x|v}(0|1) = 1 - b$ and thus this additional channel is given as $W_D(1-a,1-b)$. The modified input channel probability is denoted as $q_x^+$. Fig. \ref{fig:EveModel}(a) presents the most general broadcast wiretap channel block diagram as proposed in \cite{Csiszar1978} and Fig. \ref{fig:EveModel}(b) shows the cascade channel model we just described for Eve when the channel degradation condition does not hold (wiretap channel is less noisy), the detailed notation for the transition probabilities is shown in Fig. \ref{fig:EveModel}(c).

\subsection{Detection scenarios}
In the next sections, we obtain the secrecy capacity for different detection scenarios:
\begin{enumerate}
\item Bob and Eve use optimal quantum detection \textbf{(we denote this detection scenario as QQ)}.
\item Bob uses optimal classical detection and Eve uses optimal quantum detection \textbf{(we denote this detection scenario as CQ)}.
\item Bob uses optimal classical detection while Eve is only limited by quantum mechanics \textbf{(we denote this cryptographic scenario as DW)}.
\end{enumerate}

We denote as "classical" detection for OOK single photon detection and for BPSK coherent (homodyne or heterodyne) detection\footnote{While both photon detection and coherent detection do make use of quantum photoelectric properties, their statistics can be explained with classical electrodynamics and they are also limited by quantum noise. For this reason we denote these detectors as classical (also known as "semi-classical" or "quantum-limited"), thus highlighting the distinction with quantum detection, which beats the limit imposed by quantum noise.}. Note that case case 3) represents the strongest cryptographic scenario as it considers Eve as powerful as allowed by quantum mechanics hence not limited by detection technology. In this case, we use the Devetak-Winter rate \cite{DW2005} as the theoretical limit of achievable rate, which is defined as 
\begin{align}
C_{DW} (\gamma)
& =  \max_{V} I(V:Y)  - \mathcal{\chi}(V:E|\gamma),
\end{align}
where $E$ is the quantum system at Eve and the quantity $\mathcal{\chi}(V;E|\gamma)$ is the Holevo bound for the eavesdropper \cite{Holevo1973}. The block diagram of this case is illustrated in Fig. \ref{fig:EveModel} (d).

\section{Secrecy capacity for OOK}
Receivers with detection of coherent states beyond the standard quantum limit (SQL) are called “quantum receivers” and  have  been  extensively  studied. The first design to achieve Helstrom performance to discriminate two states was the design by Kennedy in \cite{Kennedy1973}, which achieves the minimum error probability in the high-photon-number regime. The Dolinar receiver in \cite{Dolinar1973} achieves the Helstrom limit and has been proved experimentally with feedback in \cite{Cook2007} while hybrid approaches without feedback \cite{Takeoka2005,Takeoka2008} have also been experimentally demonstrated \cite{Wittmann2008,Tsujino2008,Tsujino2011}. In the following, we obtain the binary secrecy capacity for our three detection scenarios.
\subsection{Detection scenario, QQ} 
The optimal quantum hypothesis-testing is given by Helstrom detection \cite[Chapter~7]{Cariolaro2015}, whereby both Bob and Eve's optical channels become binary symmetric channels (BSCs). In this case, Bob and Eve's channels are denoted as $W_B(\epsilon)$ and $W_E(\epsilon(\gamma))$ with $\epsilon = \frac{1}{2} \left(1-\sqrt{1- e^{-2 \eta N_t} } \right)$ and $\epsilon(\gamma) = \frac{1}{2} \left(1-\sqrt{1- e^{-2\gamma \eta N_t} } \right)$, respectively. Hence, the resulting wiretap channel is more capable and decoy states will not improve the performance of the protocol. Its secrecy capacity is achieved by uniform input probability, $q_0  = q_1 = 0.5$, and can be easily obtained according to (\ref{Cs_BSCs}) as
\begin{align}
C^{QQ}_s(\gamma)= h(\epsilon(\gamma)) - h(\epsilon). 
\end{align}
Therefore, the secrecy capacity in this case is positive as long as the degradation condition holds, i.e. $\gamma < 1$.

\subsection{Detection scenario, CQ}  
In this case we assume Bob uses standard single photon detectors, i.e. a threshold detector, which is affected by the dark count probability (the probability of detecting background events), denoted as $p_{dark}$ and the average number of noise photons (for a given collection angle and frequency/temporal processing windows) arriving at the detector, which we denote as $\Delta$. Usually $p_{dark}$ is quite low and does not influence much, however, the external noise $\Delta$ does influence the channel properties. Then, Bob's channel is given as $W_B(\epsilon_0,  \epsilon_1)$ with 
\begin{align}
\epsilon_0&=(1-p_{dark})e^{-\Delta},\\
\epsilon_1&=(1-p_{dark})e^{-(2\eta N_t +\Delta)}
\end{align}
In this case Eve is assumed to use quantum detection and the degradation conditions (10)-(11) do not hold for certain values of $\gamma$ and hence, we compute the secrecy capacity considering decoy states. \textbf{The intuition of why decoy states are needed} in this detection scenario can be understood as follows. In this detection scenario, Bob is at a technological disadvantage with respect to Eve, hence, the inclusion of decoy states will modify the statistics of the channel. This modification confuses Eve as much as possible, leading to the lowest possible leaked information towards Eve for the given physical channel. This improvement is captured in our formulation by the parameter $\gamma$, which will become as high as allowed by the physics of the channel, thus improving the secrecy capacity with respect to the case without decoy states.

We now compute the secrecy capacity considering decoy states. Eve's channel is denoted as $W^+_E(\gamma)=W_E(\epsilon(\gamma,q^+_x))$ with $$\epsilon(\gamma,q^+_x)=0.5\left(1-\sqrt{1 - 4 q^+_x (1-q^+_x) e^{-2 \eta \gamma Nt}}\right).$$  Let's denote $\beta^y_{00}=p_{y|v}(0|0)$ and $\beta^y_{01}=p_{y|v}(0|1)$ then for Bob's channel we have
\begin{align}
\beta^y_{00} &= (1-a)\epsilon_0 + a \epsilon_1,\\
\beta^y_{01} &= (1-b)\epsilon_0 + b \epsilon_1,
\end{align}
and for Eve's channel, let's denote $\beta^{z+}_{00}=p_{z|v}(0|0)$ and $\beta^{z+}_{01}=p_{z|v}(0|1)$, then
\begin{align}
\beta^{z+}_{00}(\gamma) &= (1-a)(1-\epsilon(\gamma,q^+_x) + a \epsilon(\gamma,q^+_x)),\\
\beta^{z+}_{01}(\gamma)  &= (1-b)(1-\epsilon(\gamma,q^+_x) + b  \epsilon(\gamma,q^+_x)).
\end{align}
The resulting binary non symmetric wiretap channels has a non uniform input distribution and also $q^+_x = (1-a)q+(1-q)(1-b)$. Hence, the secrecy capacity is given as
\begin{align}
C^{CQ}_s(\gamma) &= \max_{a,b,q_v}  I(V:Y) - I(V:Z | \gamma),\label{eq:CCQ}
\end{align}
where
\begin{align}
I(V:Y) &=  h(\alpha^y) - q_0h(\beta^y_{00}) - q_1h(\beta^y_{01})\\
I(V:Z | \gamma) &=  h(\alpha^z(\gamma)) - q_0h(\beta^{z+}_{00}(\gamma)) - q_1h(\beta^{z+}_{01}(\gamma)),
\end{align}
with $\alpha^y =  q_0\beta^y_{00} +  q_1\beta^y_{01}$ and $\alpha^{z+}(\gamma)  = q_0\beta^{z+}_{00}(\gamma) +  q_1\beta^{z+}_{01}(\gamma)$.

\begin{figure}[tbh]
\centering
\includegraphics[scale=0.19]{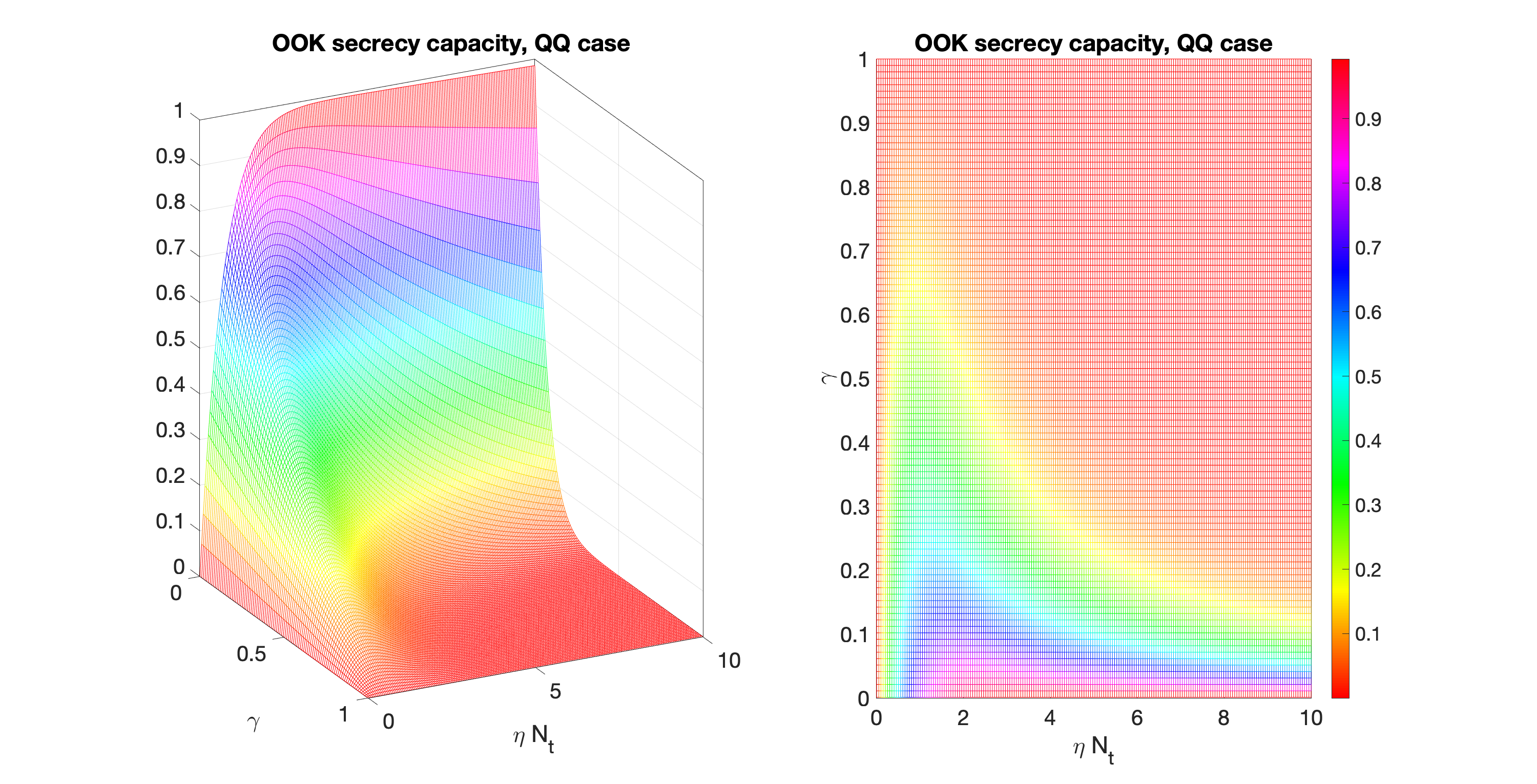}
\protect\caption{Secrecy capacity for OOK and detection scenario QQ (both Bob and Eve apply optimal quantum detection) as a function of $\gamma$ and number of arriving photons, $\eta N_t$. This wiretap channel is more capable.}
\label{fig:OOK_QQ_mesh}
\end{figure}
\begin{figure}[tbh]
\centering
\includegraphics[scale=0.13]{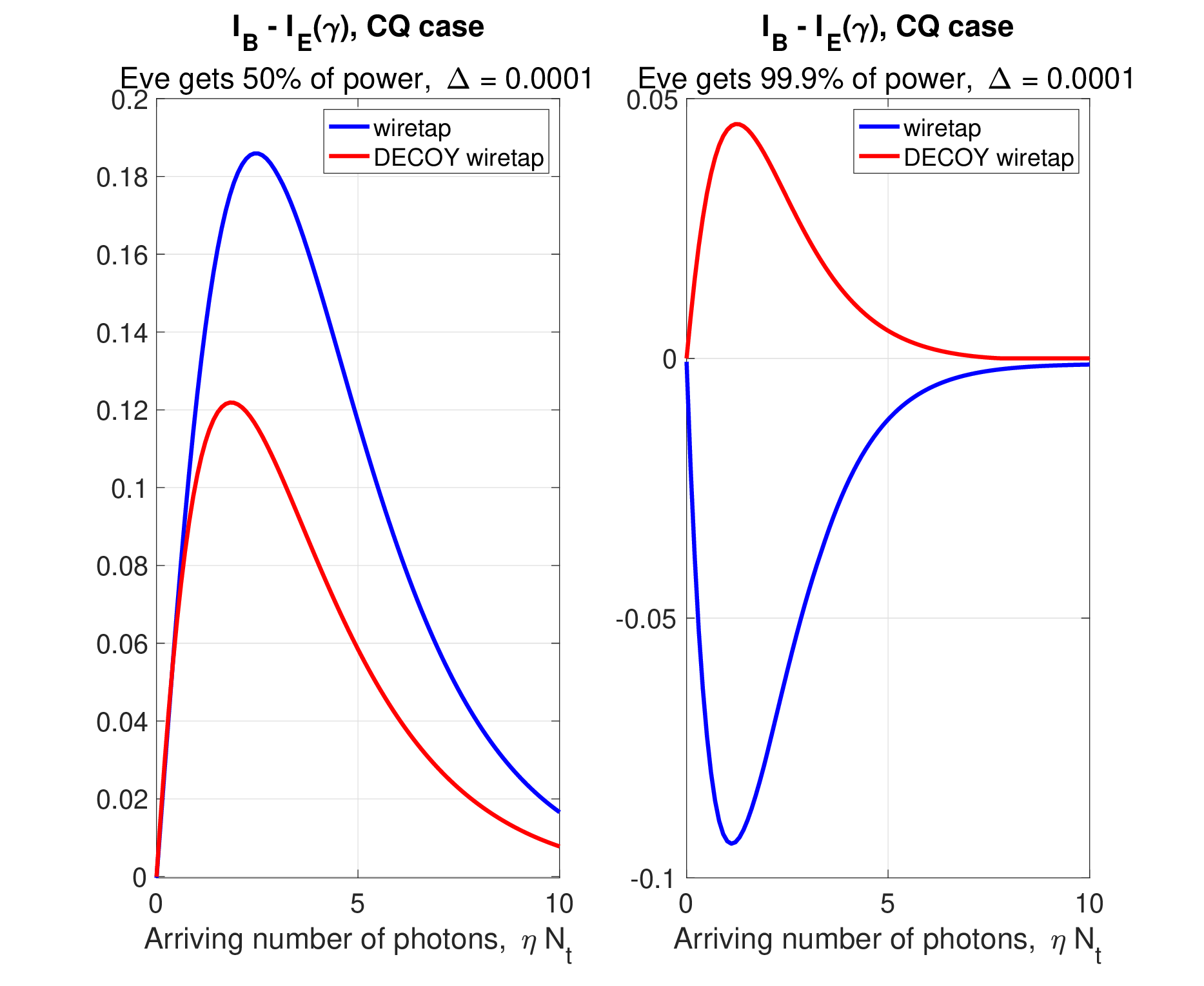}
\includegraphics[scale=0.13]{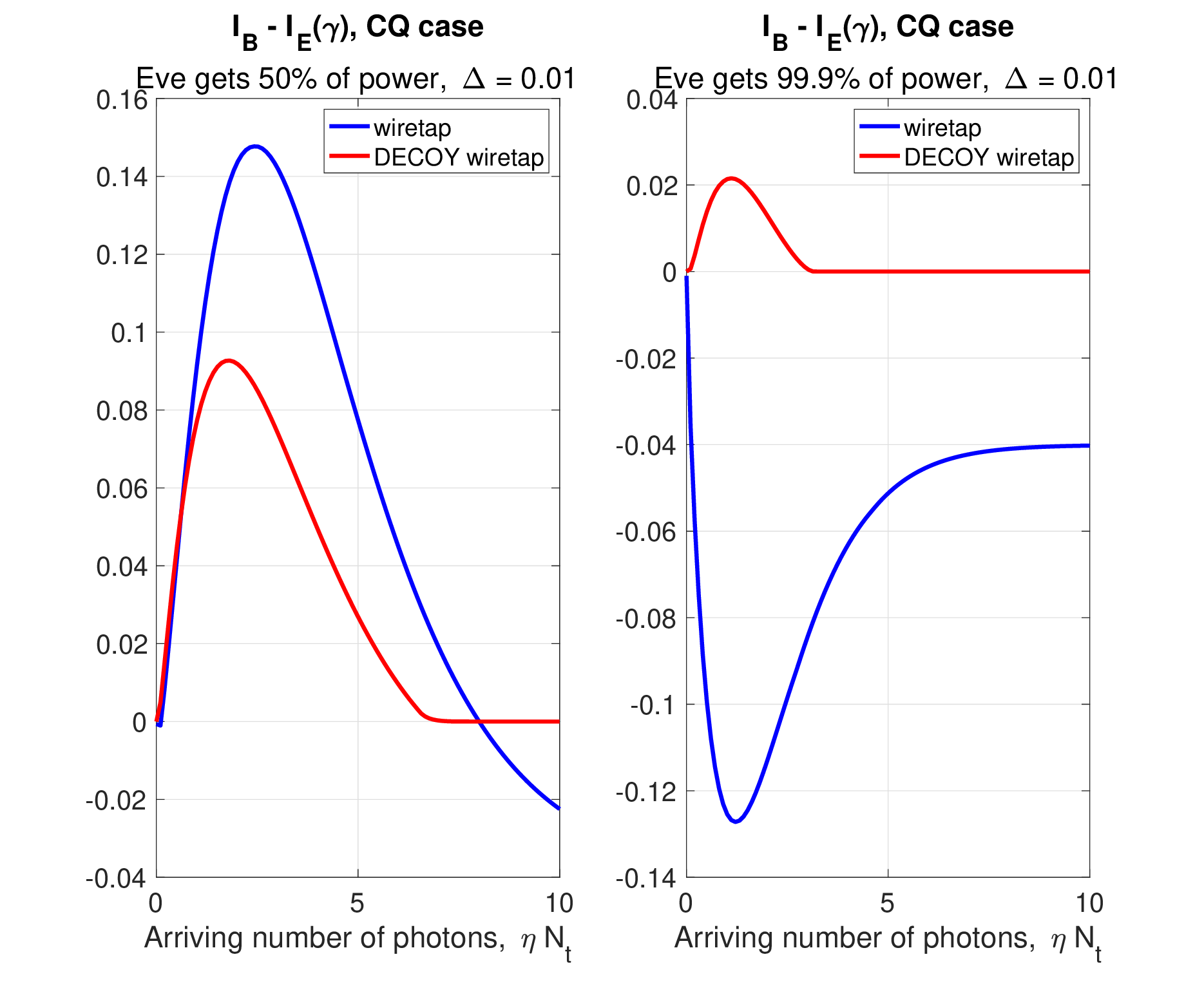}
\protect\caption{Analysis of the quantity  $I_B - I_E(\gamma)$ for the detection scenario CQ for OOK comparing two different numbers of noise photons received per pulse $\Delta = 0.0001$ (left) and  $\Delta = 0.01$ (right) as a function of the average number of photons arriving to Bob's receiver. For each case, we show results for two values of Bob’s energy fraction gathered by Eve, $\gamma = 0.5$ (left) and $\gamma=0.999$ (right)}.
\label{fig:OOK_CQ}
\end{figure}
\begin{figure*}[!ht]
    \centering
    \includegraphics[width=18.5 cm]{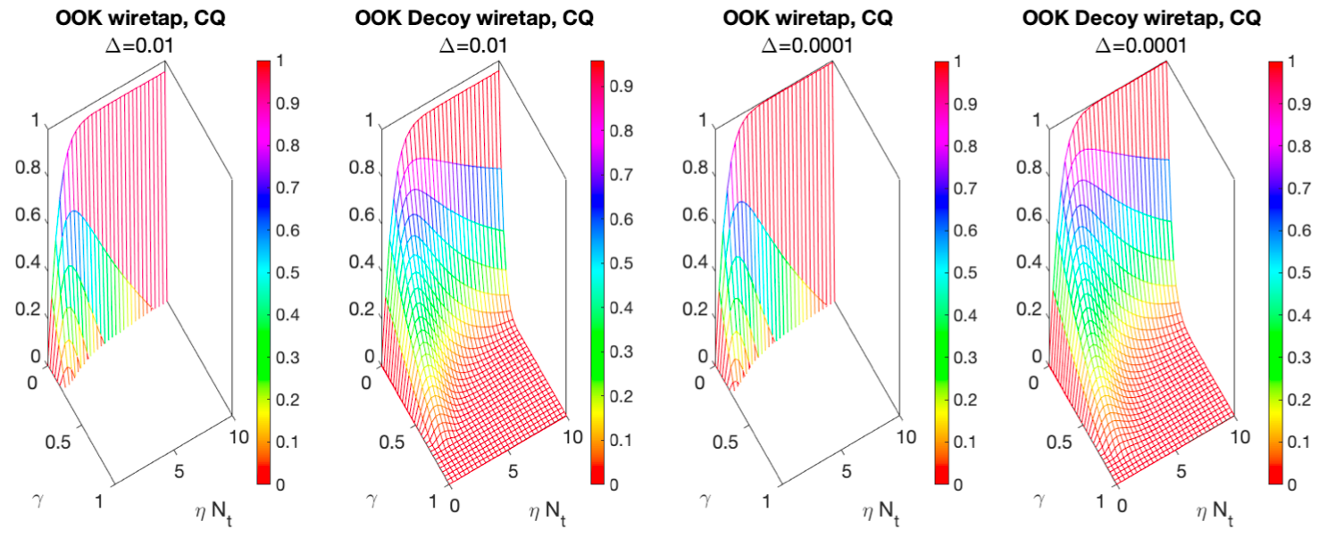}
    \caption{Secrecy capacity for OOK and detection scenario CQ as a function of $\gamma$ and number of arriving photons, $\eta N_t$ for two different external noises to Bob's photon counting detection of $\Delta = 0.01$ (left) and $\Delta = 0.0001$ (right). This wiretap channel is less noisy and the decoy states allow positive secrecy rates up to $\gamma=0.999$.}
    \label{fig:OOK_CQ_mesh_four_row}
\end{figure*}

\subsection{Cryptographic scenario, DW}  

Here we assume Bob uses photon counting detection and thus the same expressions apply with conditional probabilities  $\epsilon_0$ and $\epsilon_1$ as previously defined, while Eve is only limited by quantum mechanics. In this case the maximum rate is given as
\begin{align}
C^{DW}_s(\gamma) &= \max_{a,b,q_v}  I(V:Y) - \mathcal{\chi}(V;E|\gamma),
\end{align}
where the Holevo bound \cite{Holevo1973}\cite{Angeles2021} is given as $\mathcal{\chi}(X;E|\gamma) = 0.5(1+e^{-\gamma \eta N_t})$. 

\subsection{Numerical results}  
For the numerical results shown in this section, we solve numerically the secrecy capacities (and thus the optimal decoy probabilities) for each detection scenario, obtained in the previous sections. For the first detection scenario, the QQ scenario, Fig. \ref{fig:OOK_QQ_mesh} shows the secrecy capacity as a function of $\gamma$ and average number of photons at Bob's detector, $\eta N_t$. We observe the secrecy capacity decreases as $\gamma$ and the received number of photons increase. We also observe that the optimal received number of photons is around 1 photon.

\begin{figure}[tbh]
\centering
\includegraphics[scale=0.24]{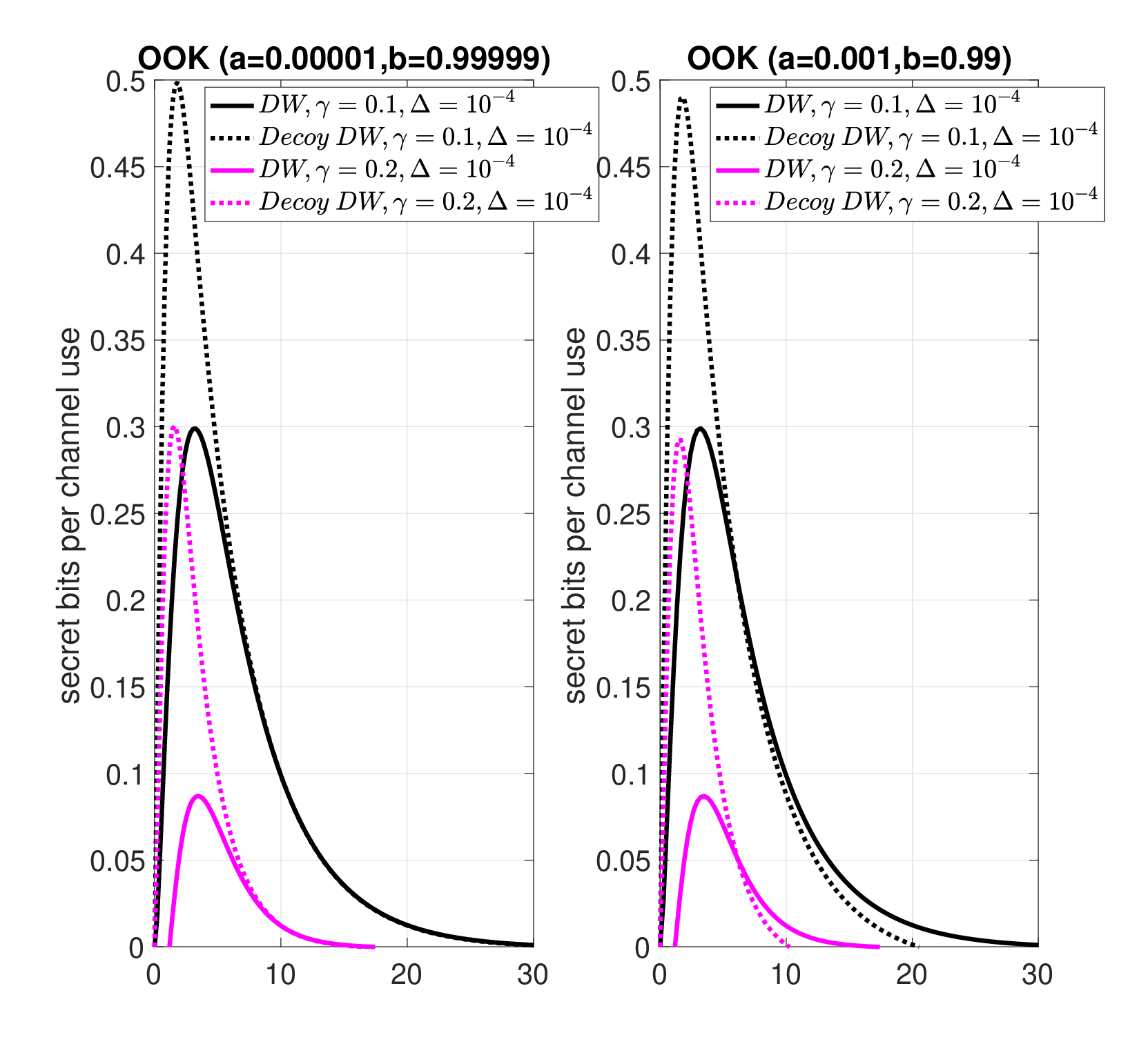}
\protect\caption{Secrecy capacity for OOK and detection scenario DW as a function of $\gamma$ and number of arriving photons, $\eta N_t$ for different Decoy probabilities.}
\label{fig:OOK_DW}
\end{figure}

Fig. \ref{fig:OOK_CQ} shows two numerical illustrative examples of the CQ case. It is assumed the single photon detection process at Bob's receiver is affected by external noise (e.g. stray light). Specifically, Fig. \ref{fig:OOK_CQ} shows the cases for $\gamma = 0.5$ and $\gamma = 0.999$  for two different external noises to Bob's photon counting detection of $\Delta = 0.00001$ (left) and  $\Delta = 0.01$ (right). We have plotted the quantity $I_B - I_E(\gamma)$ to visualize that, without decoy states, it is positive for small values of $\gamma$ while as $\gamma$ increases it becomes negative. However, with decoy states, the quantity $I_B - I_E(\gamma)$ becomes positive, thus making it possible to guarantee information-theoretic security. This is so even when Eve gathers up to 99.9\% photonic energy in the two cases of low external noise (e.g. during the night) but also with significant external noise (e.g. during the day). Interestingly, the optimal value for the highest values of $\gamma$ again occurs at around 1 photon. \textbf{Our intuition of this interesting result is that} we are assuming single-photon detection and therefore, the optimization finds that there is no point in receiving more photons, in which case they could be leaked to Eve. The secrecy capacity for the CQ case is shown in Fig. \ref{fig:OOK_CQ_mesh_four_row}, showing how decoy states allow positive rates up to $\gamma=0.999\%$. Finally, Fig. \ref{fig:OOK_DW} shows the secrecy capacity for OOK and detection scenario DW. In this case, we show that the probability of decoy states can be adjusted without significantly impacting the secrecy capacity.  Note that to fully characterize the range of decoy probabilities we can set for protocol design, a sensitivity analysis would be required, which we leave out for future work.

\section{Secrecy capacity for BPSK}
Of the available optical technologies, homodyne BPSK has a number of merits, such as frequency filtering by phase locking loop (far more selective than available optical coatings) and this very effectively discards unwanted noise and even allows to maintain a communication link if the Sun is within the receivers field-of-view. 
\subsection{Detection scenario, QQ}
As for OOK, the smallest physically allowed error probability between transmitted non-orthogonal states is given by the Helstrom bound, which in this case induces the BSCs for Bob $W_B(\epsilon_{co})$ with $\epsilon_{co} = \frac{1}{2} \left(1-\sqrt{1- e^{-4 \eta N_t}}\right)$ and for Eve $W_E(\epsilon_{co}(\gamma))$ with $\epsilon_{co}(\gamma) = \frac{1}{2} \left(1-\sqrt{1- e^{-4 \gamma \eta N_t}}\right)$, for Bob and Eve respectively. Hence,  
\begin{align}
C^{QQ}_{co}(\gamma)= h(\epsilon_{co}(\gamma)) - h(\epsilon_{co}).
\end{align}
Note that since this channel is "more capable", the secrecy capacity is positive as long as the degradation condition holds, i.e. $\gamma < 1$. \\

\subsection{Detection scenario, CQ} 
While the quantum optimal receiver can largely surpass the coherent (homodyne/heterodyne) detection limit, it is still a practical assumption for Bob's detection the use of today's technology, specially for space channels since such technology is already space-proof available. In this case, the output is Gaussian distributed \cite[Chapter~13]{Cariolaro2015}. Hence, the natural upper bound for Bob's detector is the Gaussian Shannon capacities with continuous Gaussian input. On the other hand, a useful detection lower bound is to assume that Bob uses a simple hard detection, which again induces a BSC in this case with error probabilities $P_{e}^{co} =0.5 \erfc(\sqrt{2 \eta N_t})$ for Bob \cite{Castro2023}. As for OOK, since Eve is assumed to use quantum detection the degradation conditions (10)-(11) do not hold for certain values of $\gamma$ and hence, we compute the secrecy capacity considering decoy states.

Here, we analyze the worst case scenario for Bob: the lower detection Gaussian bound. Let's denote $p_{y|x}(0|0) = 1 - P_e^{co}$ and  $p_{y|x}(0|1) = P_e^{co}$, and $\beta^y_{00}=p_{y|v}(0|0)$ and $\beta^y_{01}=p_{y|v}(0|1)$ then for Bob's channel we have
\begin{align}
\beta^y_{00} &= (1-a)(1 - P_e^{co}) + a P_e^{co},\\
\beta^y_{01} &= (1-b)(1 - P_e^{co}) + b P_e^{co},
\end{align}
and for Eve's channel
\begin{align}
\beta^{z+}_{00}(\gamma) &= (1-a)(1-\epsilon_{co}(\gamma,q^+_x) + a \epsilon_{co}(\gamma,q^+_x),\\
\beta^{z+}_{01}(\gamma)  &= (1-b)(1-\epsilon_{co}(\gamma,q^+_x) + b \epsilon_{co}(\gamma,q^+_x).
\end{align}
The resulting
secrecy capacity is given by expression (\ref{eq:CCQ}).

\subsection{Cryptographic scenario, DW}  

Here we assume Bob uses homodyne detection and thus the same expressions apply as in the previous case while Eve is only limited by quantum mechanics. In this case the maximum rate is given as
\begin{align}
C^{DW}_s(\gamma) &= \max_{a,b,q_v}  I(V:Y) - \mathcal{\chi}(V;E|\gamma),
\end{align}
where the Holevo bound \cite{Holevo1973}\cite{Angeles2021} is given as $\mathcal{\chi}(X;E|\gamma) = 0.5(1+e^{-2\gamma \eta N_t})$.

\subsection{Numerical results}
\begin{figure}[tbh]
\centering
\includegraphics[scale=0.09]{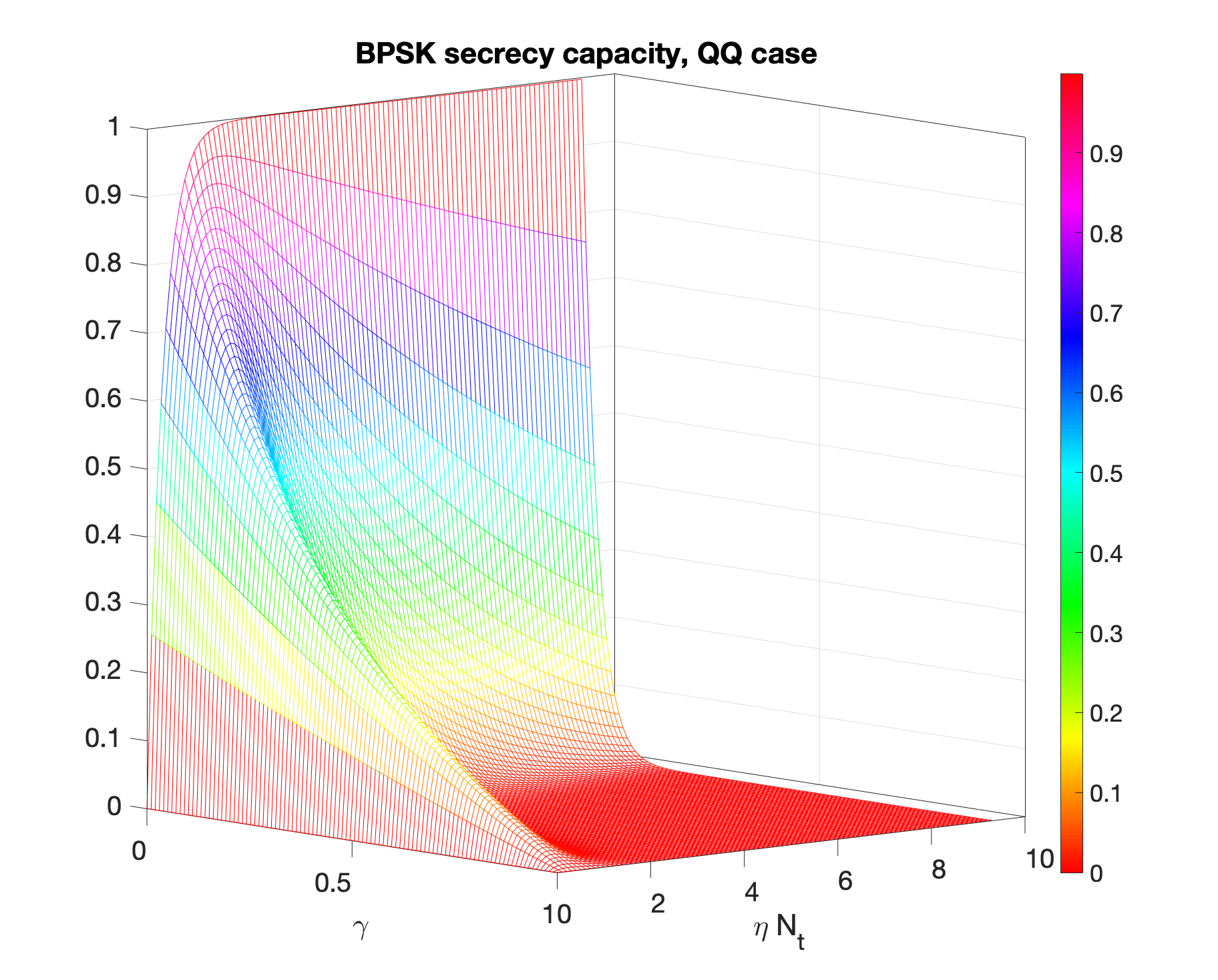}
\includegraphics[scale=0.19]{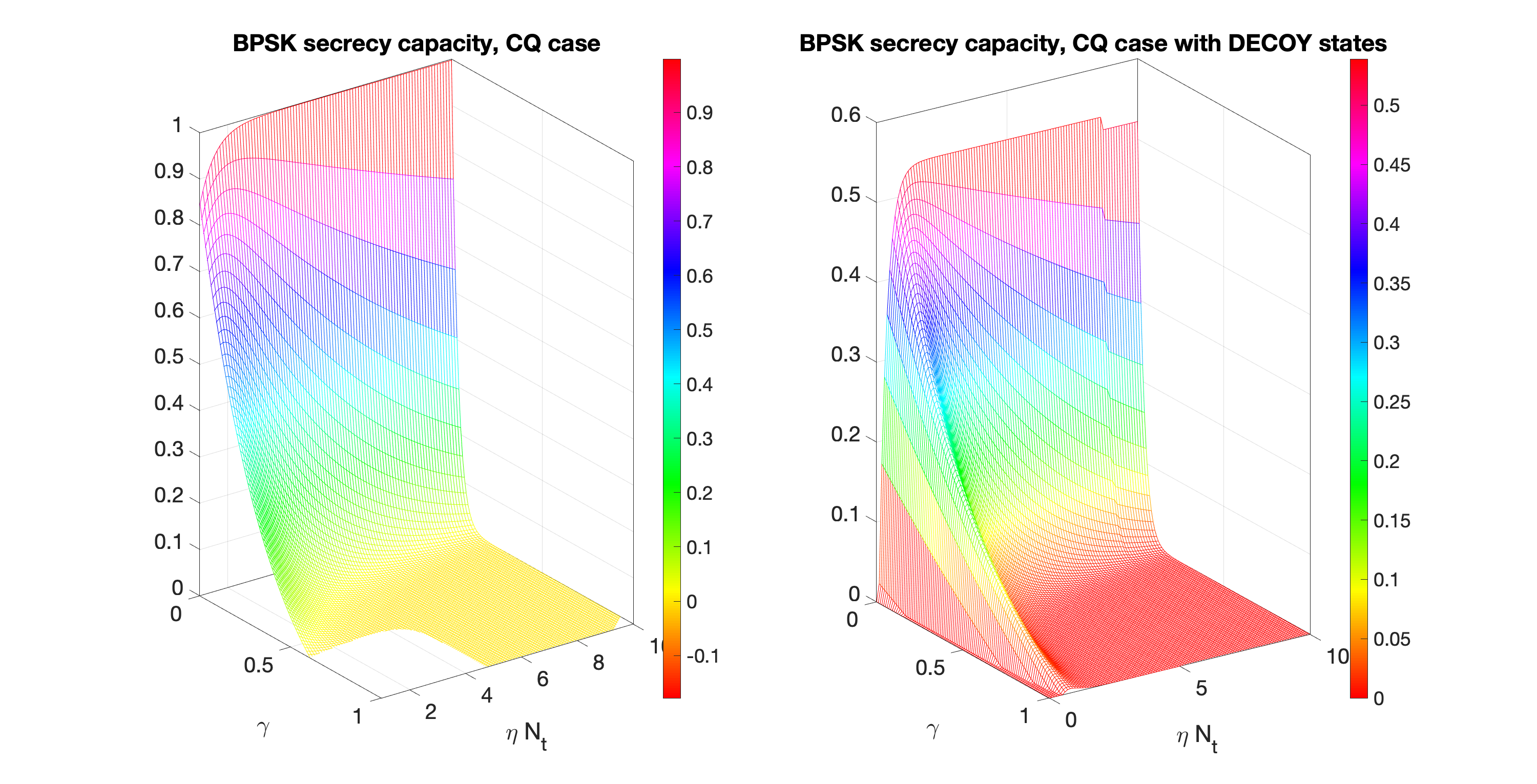}
\protect\caption{Top: Secrecy capacity for BPSK for detection scenario QQ. Bottom: Secrecy capacity for BPSK for detection scenario CQ without decoy states (left) and with decoy states (right). }
\label{fig:BPSKQQyCQ}
\end{figure}

\begin{figure}[tbh]
\centering
\includegraphics[scale=0.3]{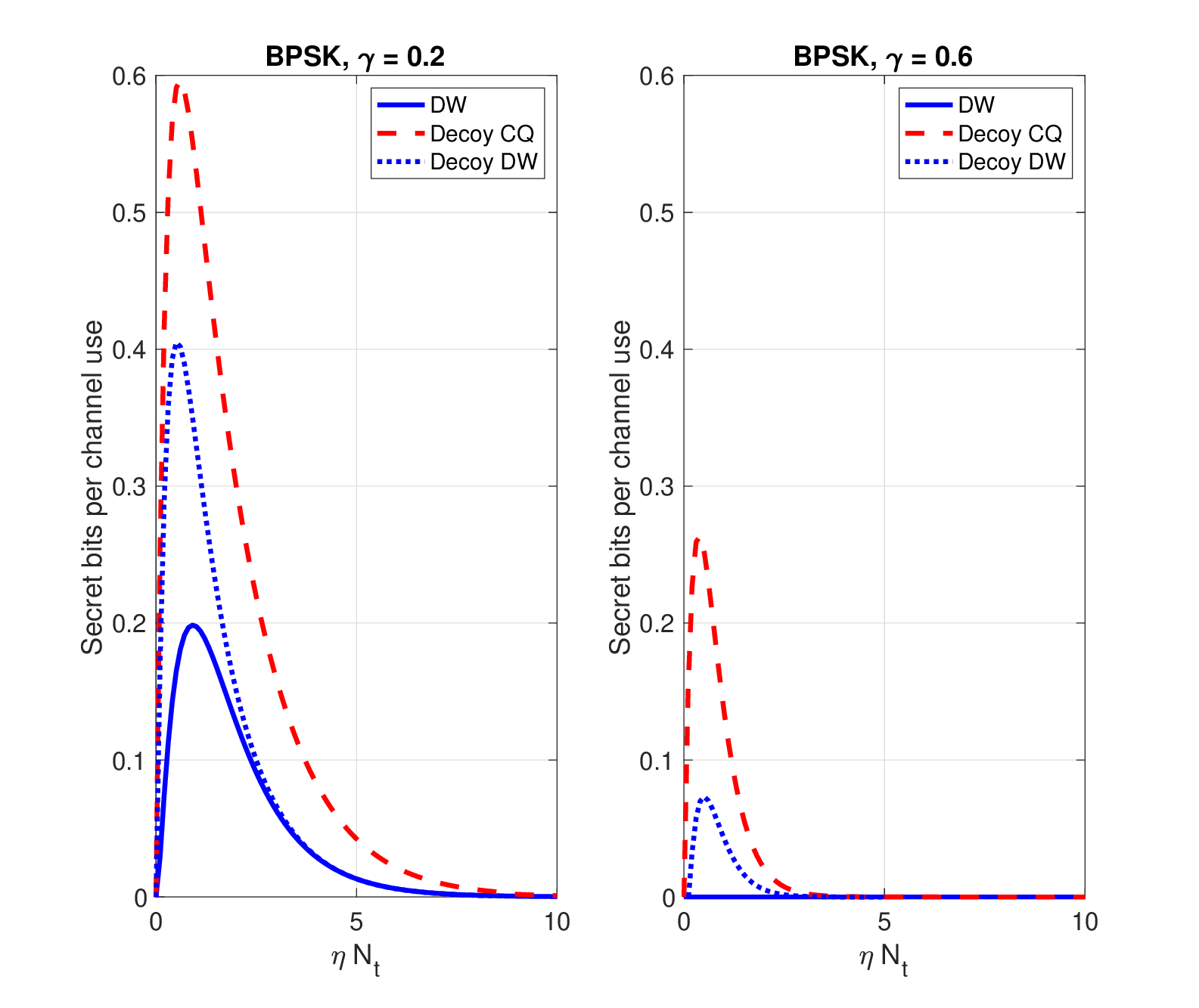}
\protect\caption{BPSK secrecy capacity for DW case without decoy states compared with DW case with decoy states and with CQ case with decoy states for two different values of $\gamma$. }
\label{fig:BPSKcomparisonDW}
\end{figure}

For the first detection scenario, the QQ scenario, Fig. \ref{fig:BPSKQQyCQ}(top) shows the secrecy capacity as a function of $\gamma$ and average number of photons at Bob's detector, $\eta N_t$. As with OOK, the secrecy capacity decreases as $\gamma$ and the received number of photons increase. We also observe that the optimal received number of photons is also around 1 photon. Fig. \ref{fig:BPSKQQyCQ}(down) shows the more interesting CQ scenario. We observe that without decoy states the secrecy capacity is zero above $\gamma=0.6$. However, it becomes positive when using decoy states for up to $\gamma = 0.999$, being again the optimal received number of photons around 1 photon. Note that using decoy states with BPSK shows lower capacity for smaller $\gamma$ values than without decoy states, resulting in an operational trade-off that is not present in OOK. Of course, the higher the $\gamma$ the lower the secrecy capacity, but it is remarkable that remains positive. Especifically, the secrecy capacity for the CQ case is  $0.6$, $0.2$, $0.1$  secret bits per channel use for values of $\gamma$ of $0.2$, $0.7$, $0.99$, respectively. Finally we analyze the DW case in comparison with CQ. This is shown in Fig.  \ref{fig:BPSKcomparisonDW} where we observe the same behaviour as with OOK, the decoy states increase the secrecy capacity for the DW scenario ensuring positive secrecy capacity even for $\gamma=0.8$. Of course, also here the higher the $\gamma$ the lower the secrecy capacity, but it is even more remarkable that remains positive for the DW assumptions. Especifically, the secrecy capacity for the DW case is  $0.41$, $0.02$, $0.006$  secret bits per channel use for values of $\gamma$ of $0.2$, $0.7$, $0.8$, respectively. From such a better performance than OOK for the DW scenario we can conclude that our protocol provides better secrecy guarantees for BPSK.

%

\section{Design methodology for the inter-satellite channel study case}
\begin{figure}[tbh]
\centering
\includegraphics[scale=0.12]{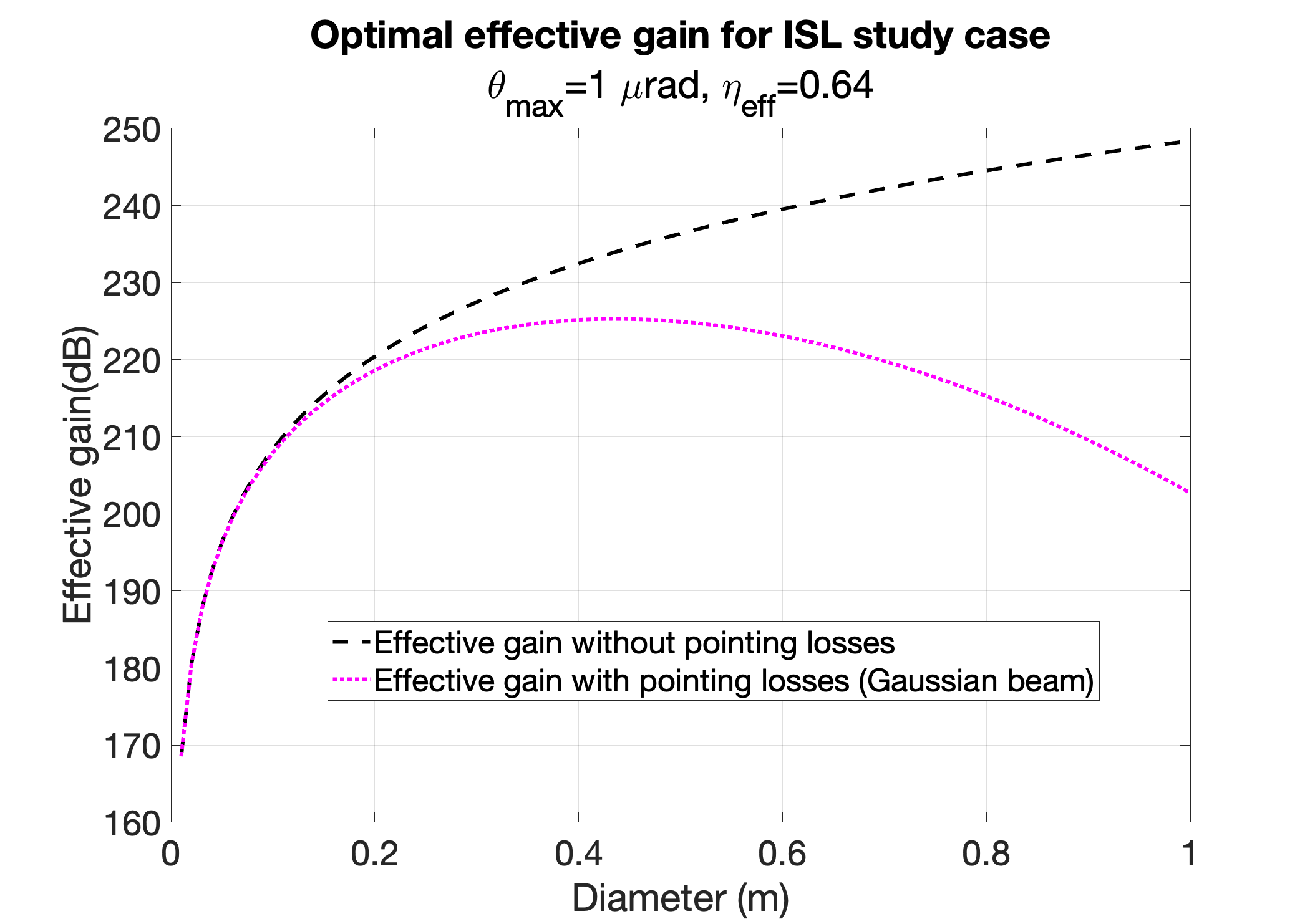}
\includegraphics[scale=0.15]{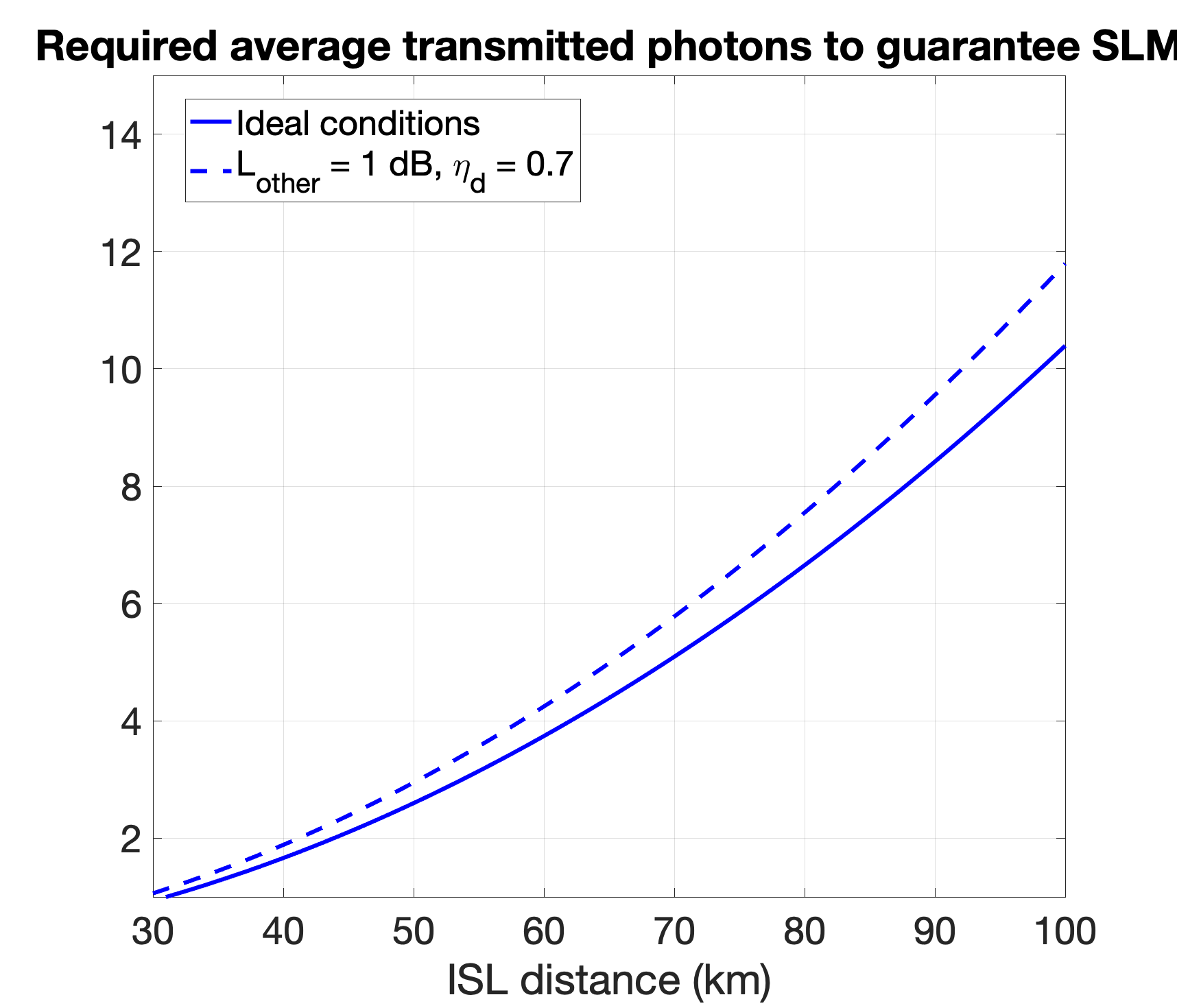}
\protect\caption{Top: Optimal gain is shown to be $G^*_{eff}=225$ [dB] for telescopes of 40 cm diameter at transmission and reception. Bottom: required number of transmitted photons to guarantee the security margin for $N^*_B$=1. Note that e.g. 10 photons per pulse correspond to only 6,4 nW for the parameters of Table 1. We also note that $\gamma$ helps define the actual link budget of the protocol to achieve the optimal operational point (i.e. that the secrecy capacity is maximum) while allowing for secured leakage towards Eve, in this calculations we have considered $\gamma=1$. }
\label{fig:OptimalGain}
\end{figure}

We now consider the study case of inter-satellite links (ISLs) that establish satellite-to-satellite communication in intra- or inter-orbital links. To enable high-speed data transfer over the ISL, the optics of the transmitter and receiver must first establish a mutual line of sight tracking. This requires a tracking accuracy of about 1 $\mu rad$ \cite{Chen1989} because of the narrow beamwidth of an optical ISL. In general, a closed-loop tracking servo can help achieve the desired pointing accuracy at transmission by locking onto the beacon signal from the remote receiver. The beacon signal can either be a separate (wide-beam) laser or the information-carrying signal itself. To detect the beacon signal, a quadrant detector is typically used. Along with this pointing and tracking calibration, the practical design of a secure quantum communication protocol also requires the detailed calibration of the quantum link and the protocol according to some specific target information-theoretic security and error rate, which requires finite-length analysis. However, in this work we focus on the asymptotic performance of our quantum protocol, i.e. the best performance our protocol can achieve, for which we only need to design the operational link budget according to the optimal number of photons at Bob's reception (denoted as $N^*_B$) for the target leakage allowed by our protocol to be guaranteed as secure against Eve's cryptanalysis (parameterised by $\gamma$ in this work). This is manageable in an ISL channel, where atmospheric effects are absent and therefore only pointing losses affect the link quality. For our free space ISL we have that the optical losses are given as the ratio of the telescope area and the footprint area
\begin{align}
\eta= \frac{A_R}{A_F} L_{p,tot} L_{other},
\end{align}
where $L_{p,tot}$ are the total (transmitter and receiver) pointing losses, and $L_{other}$ are implementation losses. In order to do engineering optimization, the above expression can be written as 
\begin{align}
\eta=\eta_d G_t \eta_t G_r \eta_r \left( \frac{\lambda}{4 \pi R}\right)^2 L_{p,tot} L_{other},
\end{align}
where $\eta_d$ is the detector efficiency, $G_t$ and $G_r$ are the gains at transmission and reception with $\eta_t$ and $\eta_r$ their corresponding efficiencies, $\lambda$ is the laser wavelength, $R$ is the inter-satellite distance, In case of Gaussian beam, the pointing losses are given by $L_p(\theta) = e^{-G\theta^2}$, where $G$ is the linear gain of the telescope and $\theta$ is the angular error. This error is random in nature, and therefore we can follow either a deterministic approach of simply taking a maximum angular error, $\theta_{max}$, or a probabilistic approach of assuming a given outage probability, $P_{out}$. When the two errors at transmission and reception are independent and identically distributed (i.i.d.) Gaussian with zero-mean, the resulting angular distribution is Rayleigh with parameter $\sigma_{\theta}$. 
If  for simplicity (and realistically for a given satellite constellation) we assume the same parameters $\sigma_t=\sigma_r=\sigma_{\theta}$, $G_t=G_r=G$ and $\eta_{eff}=\eta_t\eta_r$ it is easy to show that \cite{Chiani2021}
\begin{align}
P_{out}=\left(1+\frac{K}{2\sigma_{\theta}^2 (\eta_{eff}G)}\right)e^{-\frac{K}{2\sigma_{\theta}^2 \eta_{eff}}},
\end{align}
with $K = \frac{ln(10)}{10} L_p$. In either case, there is an optimal optical gain, which we obtain defining an effective optical gain as
\begin{align}
G_{eff}[dB]= 2 G \eta_{eff} [dB] + 2 L_{p,tot} [dB],
\end{align}
so that $\eta= G^*_{eff} \left( \frac{\lambda}{4 \pi R}\right)^2 L_{other}$ where $G*_{eff}$ is the optimal effective optical gain. Considering the realistic optical system parameters summarized
in Table 1 and only the deterministic pointing losses, Fig. \ref{fig:OptimalGain} (top) illustrates the optimal gain for this case considering $\theta_{max}=1\mu rad$ where we see that $G^*_{eff}=225$ dB. Now, for the formulation of our link budget, we introduce the \textbf{Security Link Margin (SLM)} where we leverage and adapt the traditional concept of link margin in wireless communications, applying it innovatively to enhance the security framework of our protocol. This margin quantifies the permissible leakage that our protocol guarantees to be information-theoreticly secure, and thus it corresponds to the value $\gamma N_B$. The operational significance of such SLM is shown in the resulting link budget expression given as
\begin{align}
N^*_B = N_B + \text{SLM.}
\end{align}
Here, \( N^*_B \) is the optimal operational point (determined from our results as the condition under which the secrecy capacity reaches its maximum). Hence, this point is achieved by carefully balancing \( N_B \) adjusted for link attenuation and initial transmission energy, \( N_t \).
\begin{align}
N^*_B  = (\eta N_t) + (\gamma \eta N_t).
\end{align}
In this equation, the quantity \( (\gamma \eta N_t)  \) specifies the SLM in average number of photons. The parameter \( \gamma \) denotes the fraction of the average number of photons that can be securely intercepted, encapsulating the protocol's designed resilience against eavesdropping. Hence, this framework permits the adjustment of the transmit average number of photons---and thus the actual \(N_B\)---in response to the defined SLM, ensuring alignment with the identified optimal operational point. Our proposed SLM integrates both operational and security considerations, and thus it is a useful tool for designing quantum communication systems by showing the optimal adjustments at transmission that directly contribute to maintaining the highest possible secrecy capacity.

\begin{table}
\centering
\caption{ISL STUDY CASE PARAMETERS}
\begin{tabular}{||c c c c||} 
 \hline
                 & Symbol & Units& Value \\ 
 \hline\hline
Wavelength & $\lambda$ & $nm$ & $1550$ \\ 
\hline
Pulse rate & $B$ & $GHz$ & $5$ \\ 
\hline
Detector efficiency & $\eta_d$ & $$ & $0.7$ \\ 
\hline
Transmitter optical eff. &  $\eta_t$ & $$ & $0.8$ \\
 \hline
Receiver optical eff. &  $\eta_r$ & $$ & $0.8$ \\
 \hline
Maximum pointing error &  $\theta_{max}$ & $\mu rad$ & $1$ \\
\hline
Other Losses &  $L_{other}$ & $dB$ & $1$ \\
\hline
\end{tabular}
\end{table}
Fig. \ref{fig:OptimalGain} (bottom) shows the required number of transmitted photons $N_t$ as a function of the ISL range to guarantee the security link margin for the values in Table I and also in the ideal conditions of no implementation losses and ideal detector.  Note that with BPSK space-proof technology our protocol is readily implementable and our numerical results indicate that for our assumed pulse rate, it can guarantee a secrecy rate of 500 Mbps (CQ scenario) without the requirement to monitor Eve or any exclusion area. Bob can detect if Eve (wherever she is) has eavesdropped the communication whenever the link departs from the calibrated received number of photons corresponding to the security link margin, \textbf{in which case the protocol aborts}. Of course, the protocol can be also implemented using $P_{out}$ to detect Eve, this case is left for further work. Moreover, we remark that our just presented novel design methodology of guaranteed optical space ISLs is independent of the pulse rate, which will define the ultimate achievable secrecy rates, which will be higher as pulse rates become higher. Further, our methodology can be carried over to the finite-length regime, whose detailed design we leave for our subsequent work.

\section{Potential gains with non-classical squeezed states}
Coherent states have circularly symmetric uncertainty regions and Poissonian statistics, just like coherent light (see Fig. \ref{fig:CohSqz}). A further reduction of uncertainty is possible by “squeezing” the uncertainty region breaking the circular symmetry of coherent states, which then allows detection with reduced quantum noise with respect to coherent states, a property that cannot be explained within classical physics. Squeezed light has already been well investigated for several applications, including quantum communication from practical \cite{Yuen1, Yuen2, Yuen3, Paris2001} and information-theoretic \cite{HolevoSohmaHirota1999} points of view, and were first produced in the mid 1980s (at that time squeezed states were also known as two-photon coherent states) using wave mixing in an optical cavity \cite{Wu1986} and parametric down conversion \cite{Dodonov1998}, but can also be obtained by photon-adding on coherent states \cite{Curado2021}. While the squeezing parameter can be any complex number, for our case and without loss of generality, we can assume $r \in \mathbb{R}$ and $r>0$. They can be expressed in terms of the Fock basis as \cite{Gong1990}, and in our case the expression is

\begin{figure}[tbh]
\centering
\includegraphics[scale=0.2]{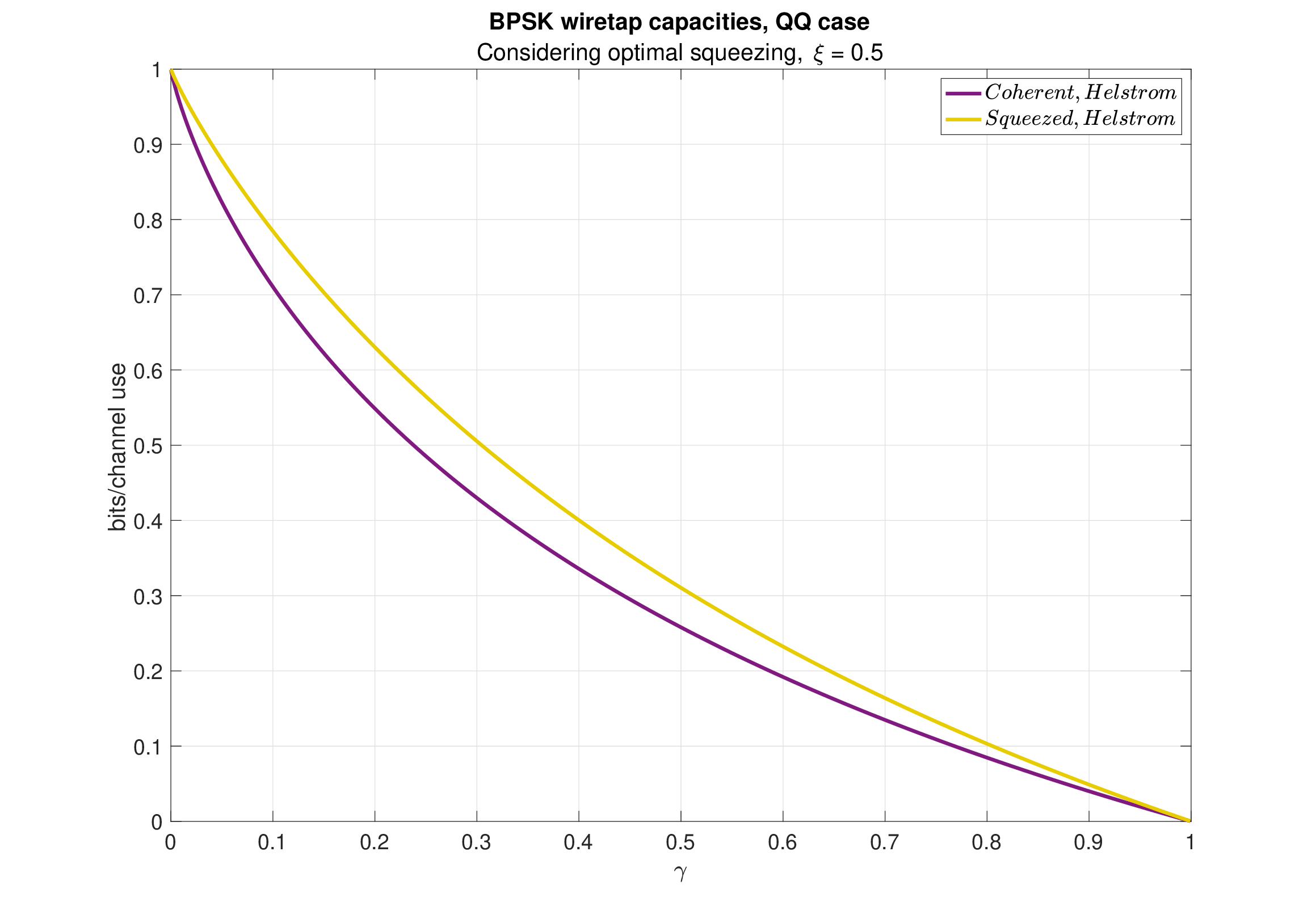}
\includegraphics[scale=0.2]{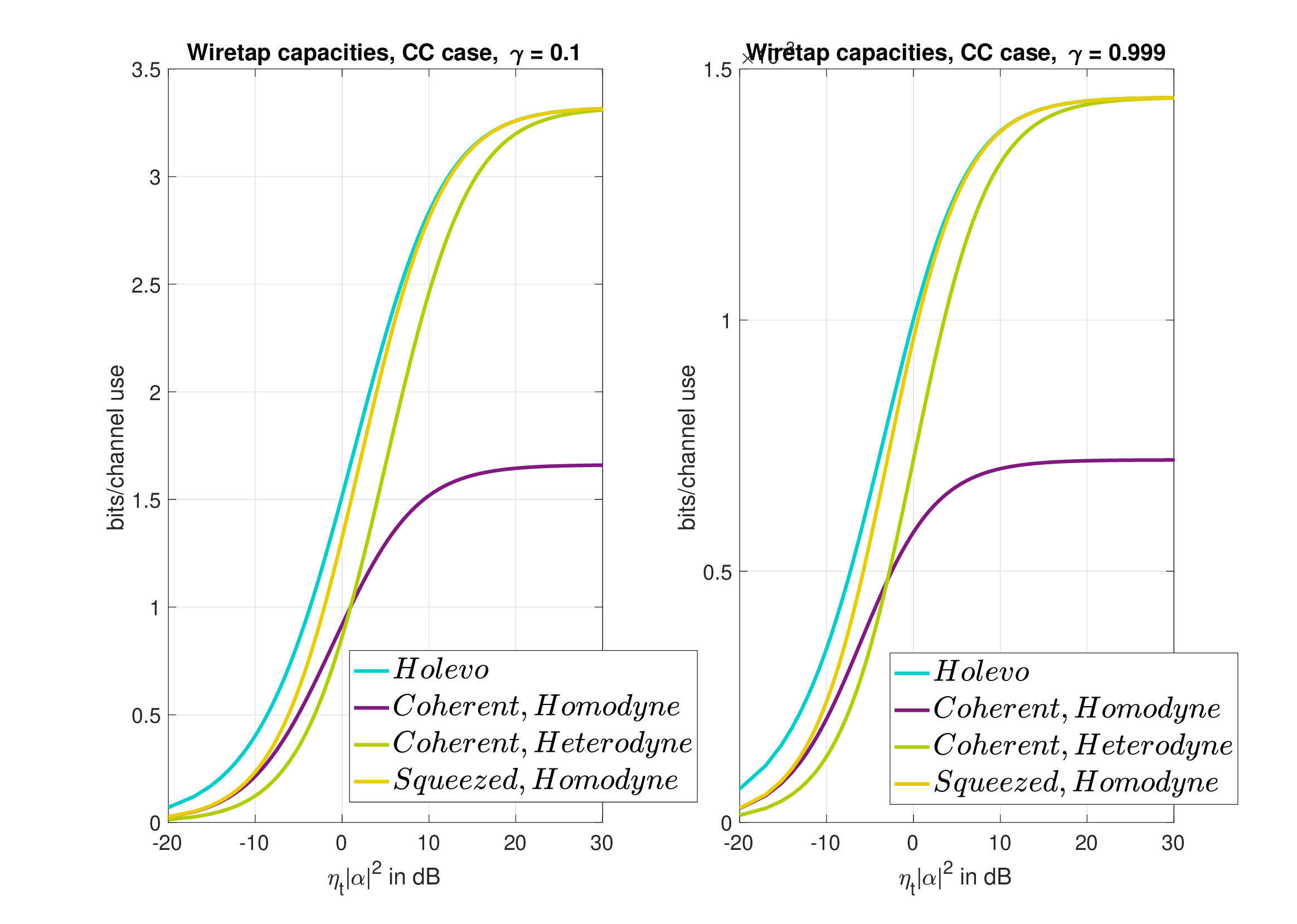}
\protect\caption{Top: Dependency with $\gamma$ of the secrecy capacity for coherent and squeezed states. Bottom: Wiretap capacities for two different values of $\gamma$ for coherent and squeezed states and Holevo bound.}
\label{fig:QQcase_Cs_gamma}
\end{figure}

\begin{figure}[tbh]
\centering
\includegraphics[scale=0.14]{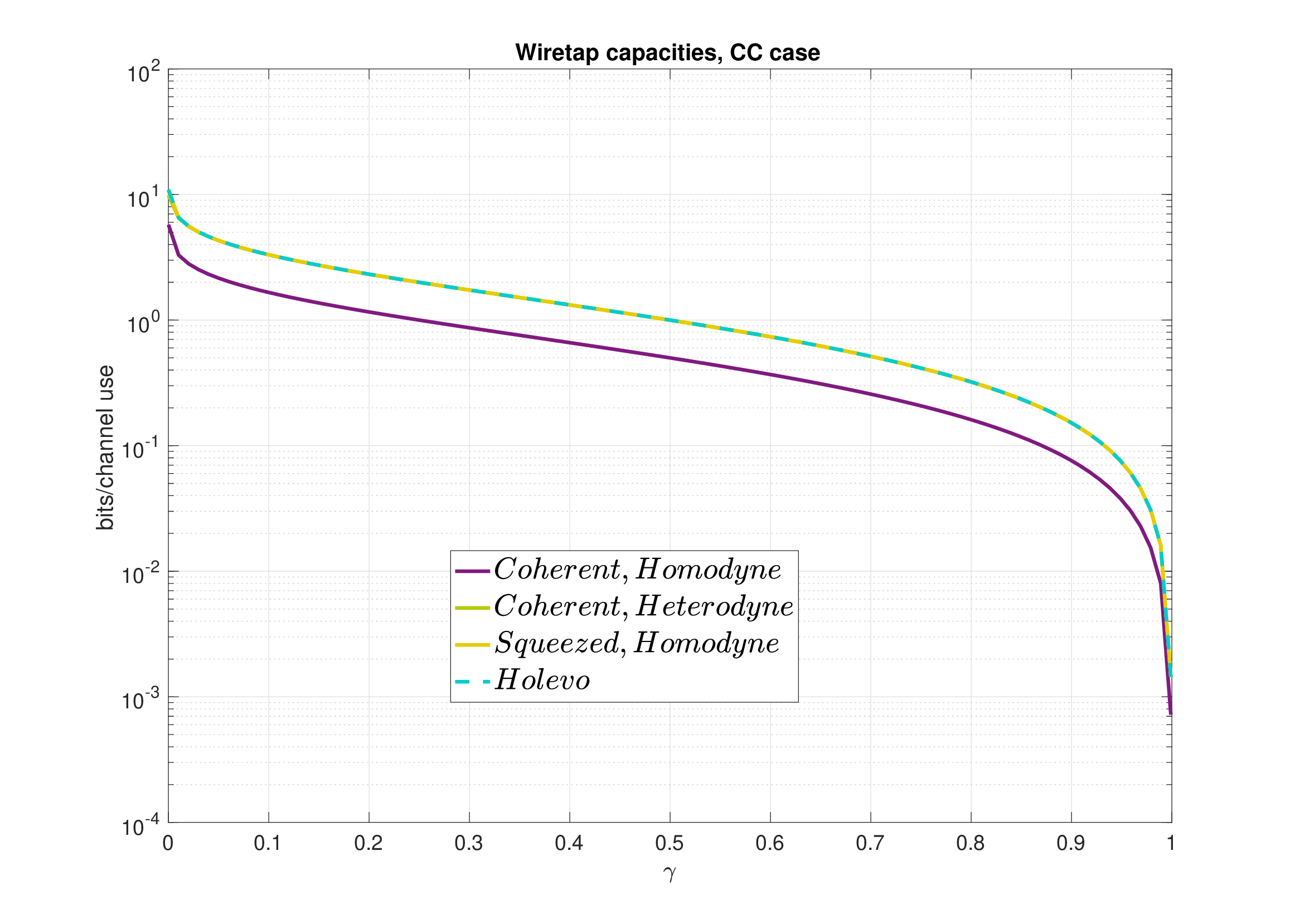}
\protect\caption{Wiretap capacities for CC case comparing coherent states and squeezed states.}
\label{fig:CCcase_Cs_gamma}
\end{figure}

{\small
\begin{align}
\ket{\psi} &= \ket{\alpha,r}= \frac{G(z,r)}{\sqrt{\cosh r}}\sum_{n=0}^{\infty} \frac{1}{\sqrt{n!}}\left( \frac{\tanh r}{2} \right)^{\frac{n}{2}}) H_n(z^*) \ket{n},
\end{align}
} 
where 
{\begin{align}
G(z,r) = \exp \left[  \frac{-|z|^2}{2} \sinh(2r) + (z^*)^2  \sinh^2(2r)   \right],
\end{align}
} 
with $z = \frac{1}{2}\alpha^*e^{\frac{j}{2}\theta} \frac{1}{\sqrt{\tanh r}} + \frac{1}{2}\alpha e^{\frac{-j}{2}\theta} \frac{1}{\sqrt{\tanh r}} $ and the $H_n( . )$ are the Hermite polynomials analytically continued to the whole complex plane. The transmitted signal is in one of two squeezed states, represented as $\psi_0 = \ket{-\alpha, r}$ and $\psi_1 = \ket{\alpha, r}$  for encoding bits $0$ or $1$, respectively for BPSK and  $\psi_0 = \ket{0, r}$ and $\psi_1 = \ket{\alpha, r}$ for OOK. We denote the average number of transmitted photons as 
\begin{align}
N_{t,sq} = q_0 |\psi_0|^2 + q_1 |\psi_1|^2, 
\end{align}
We now define the fraction of squeezing as
\begin{align}
\xi = \frac{\sinh^2 (r)}{N_t}, 
\end{align}
where $\sinh^2 (r)$ is the average photon number of a squeezed vacuum state. Hence, $\xi$ represents the ratio of the mean photon number due to squeezing to the mean photon number of a corresponding coherent state with the same initial energy. Hence, the average number of transmitted photons is 
\begin{align}
N_{t,sq} = q_0 (1+\xi)|\alpha_0|^2 + q_1 (1+\xi)|\alpha_1|^2 = (1+\xi)|\alpha|^2.
\end{align}
For a meaningful comparison of the performance when using coherent or squeezed states, we want to fix $N_t$ as the average number of transmitted photons for both coherent and squeezed states, thus having
\begin{align}
N_{t,sq} = N_t = (1 -\xi) |\alpha|^2 + \sinh^2 (r).
\end{align}
In order to decide a justified value for $\xi$ to use for our comparison, it is known \cite{Chesi2018}\cite{Yuen2} that $\xi = 0.5$ is optimal with respect to minimizing the error probability. With this value we obtain our results. Fig. \ref{fig:QQcase_Cs_gamma} (left) shows the advantage of squeezing for the QQ case, which depends on the value of $\gamma$. We see that the advantage is modest, up to 8\%. As for the CC case assuming Gaussian inputs and ideal detection, given its theoretical interest we also plot the attainable secrecy capacities in Fig. \ref{fig:QQcase_Cs_gamma} (right) together with the ultimate wiretap capacity, i.e. when Bob and Eve can attain the Holevo bound \cite{Castro2023}, remarkably, it is attained by squeezed BPSK homodyne detection and the dependency with $\gamma$ is shown in Fig. \ref{fig:CCcase_Cs_gamma}. We observe that in this theoretical case the squeezing gain is higher (up to 50\%) when compared to the QQ case with Helstrom detection (whose optimality is derived from minimizing the probability of error). Research is needed to clarify these potential gains including the technological challenges of achieving Gaussian capacities while maintaining squeezed states.
\vspace{-4 mm}
\vspace{-5 mm}
\section{Discussion and further work}
We have presented the asymptotic security analysis of a quantum keyless secure communication protocol that transmits classical information over quantum states with dummy (decoy) pulses optimally obtained to guarantee information-theoretic security. We have obtained numerical results for OOK and BPSK, and show that our protocol significantly outperforms the conventional protocol without decoy states. We also obtain that BPSK outperforms OOK and ensures positive secrecy capacity even when Eve gathers up to 99\% of the average number of photons that Bob detects while being only limited by quantum mechanics. We have also introduced a design methodology for our protocol introducing the novel concept of security link margin. This design allows Bob to detect if Eve (wherever she is) has eavesdropped the communication, in which case the protocol aborts. Our protocol can be readily implemented with quantum technology already available leveraging the development of quantum technologies motivated by the progress of QKD protocols. Moreover, homodyne detectors are already available space-proof and therefore our protocol for BPSK is readily implementable over quantum states with current technology. For OOK, our protocol can be complemented with QKD protocols by trading-off complexity and security depending on the scenario of interest or even in real time. 
We have also shown that while theoretical advantage when using squeezed states is up to 50\%, the practical advantage seems modest and furthermore this technology is not yet available. As further work we will study more scenarios and conduct sensitivity analysis of the decoy probabilities and finite-length security analysis to identify the specific parameters of the processing algorithms and forward error correction codes to ensure that the reliability and security targets are met. We also intend to analzye the extension that makes our protocol to be also robust to other quantum attacks (and not only to eavesdropping) to converge to QSDC. With this analysis, we will be able to establish the time structure of the protocol and identify safeguards that will allow the protocol to abort and activate again as needed to maintain data integrity. The finite-length analysis will also allow to compute an operational outage probability. In doing this, we also intend to analzye the extension that makes our protocol to converge to QSDC. In this case, the protocol will be also robust to other quantum attacks (and not only to eavesdropping) at the cost of being slower and more complex. Further analysis is also needed to completely characterize the use of squeezed states for our protocol.
\vspace{-1 mm}
\section{Acknowledgments}
\noindent This work has been performed under a programme of and
funded by the European Space Agency: Satellite Network of
Experts V (SatNEx V) 4000130962/20/NL/NL/FE. The view
expressed herein can in no way be taken to reflect the official
opinion of the European Space Agency.



\end{document}